\documentclass[letterpaper]{article} %
\usepackage{aaai24}  %
\usepackage{times}  %
\usepackage{helvet}  %
\usepackage{courier}  %
\usepackage[hyphens]{url}  %
\usepackage{graphicx} %
\usepackage{natbib}  %
\usepackage{caption} %
\usepackage{algorithm}
\usepackage{algorithmic}

\usepackage{newfloat}
\usepackage{listings}
\DeclareCaptionStyle{ruled}{labelfont=normalfont,labelsep=colon,strut=off} %
\floatstyle{ruled}
\newfloat{listing}{tb}{lst}{}
\floatname{listing}{Listing}
\usepackage{booktabs}

\usepackage{numprint}
\npthousandsep{,}

\usepackage{subcaption}
\usepackage{changepage} %

\usepackage{xcolor}

\definecolor{limegreen}{RGB}{50,205,50}
\definecolor{darkgreen}{RGB}{0,100,0}

\title{iDRAMA-Scored-2024: A Dataset of the Scored Social Media Platform from 2020 to 2023}

\author {
    Jay Patel\textsuperscript{\rm 1},
    Pujan Paudel\textsuperscript{\rm 2},
    Emiliano De Cristofaro\textsuperscript{\rm 3},
    Gianluca Stringhini\textsuperscript{\rm 2},
    Jeremy Blackburn\textsuperscript{\rm 1}
}
\affiliations {
    \textsuperscript{\rm 1}Binghamton University\\
    \textsuperscript{\rm 2}Boston University\\
    \textsuperscript{\rm 3}University of California Riverside\\
    jpatel67@binghamton.edu, ppaudel@bu.edu, emiliand@ucr.edu,
    gian@bu.edu, jblackbu@binghamton.edu
}

\begin{document}

\maketitle

\begin{abstract}

Online web communities often face bans for violating platform policies, encouraging their migration to alternative platforms.
This migration, however, can result in increased toxicity and unforeseen consequences on the new platform.
In recent years, researchers have collected data from many alternative platforms, indicating coordinated efforts leading to offline events, conspiracy movements, hate speech propagation, and harassment.
Thus, it becomes crucial to characterize and understand these alternative platforms.
To advance research in this direction, we collect and release a large-scale dataset from Scored -- an alternative Reddit platform that sheltered banned fringe communities, for example, c/TheDonald (a prominent right-wing community) and c/GreatAwakening (a conspiratorial community).
Over four years, we collected approximately 57M posts from Scored, with at least 58 communities identified as migrating from Reddit and over 950 communities created since the platform's inception.
Furthermore, we provide sentence embeddings of all posts in our dataset, generated through a state-of-the-art model, to further advance the field in characterizing the discussions within these communities.
We aim to provide these resources to facilitate their investigations without the need for extensive data collection and processing efforts.

\end{abstract}

\section{Introduction}

\begin{figure}[t]
\centering

\includegraphics[width=0.96\linewidth]{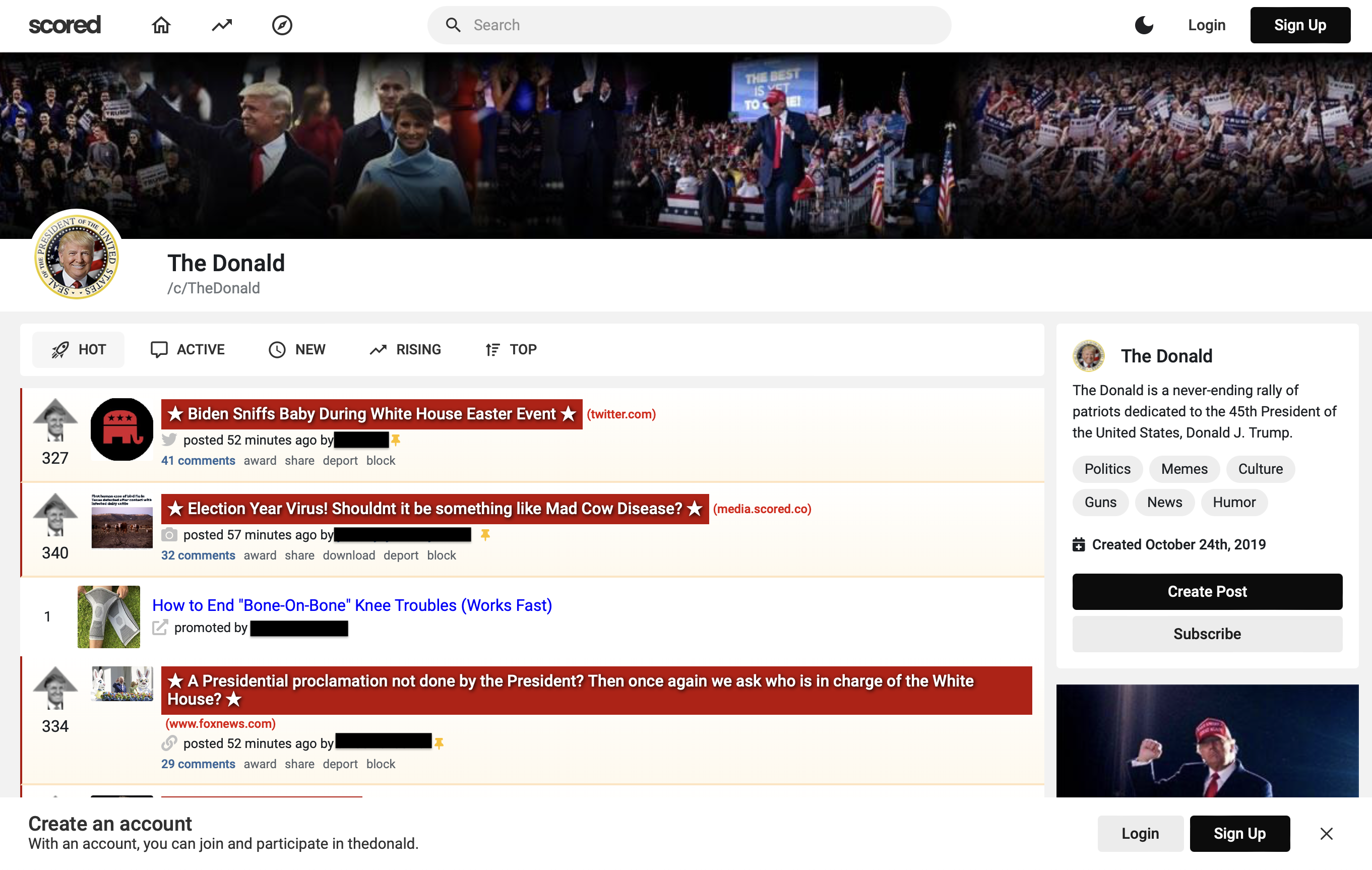}
\caption{Visual appearance of c/TheDonald's home page on Scored platform.}

\label{fig:scored_ss}
\end{figure}

Social media has become more than just a platform for social connection and entertainment, it serves as a platform for the mass consumption of news~\cite{shearer2018social}, spreading misinformation and hate speech~\cite{zannettouWebFalseInformation2019, ziems2020racism}, and propagation of political narratives~\cite{ng2022coordinated, starbird2019disinformation}.
As its use continues to rise, this has become an arms race to enhance the existing systems that facilitate a safer space for everyone.

Social platforms like Reddit offer a space for user groups with shared interests, known as communities, to share and discuss content~\cite{porter2004typology}.
However, if communities violate platform policies, Reddit imposes bans, pressing them to migrate elsewhere (e.g., alternative platforms) on the Web.
Researchers have studied various alternative platfforms, for example, Voat~\cite{papasavvaItQoincidenceExploratory2021}, Poal~\cite{papasavva2023waiting}, and Lemmygrad~\cite{balci2023data} (an alternative to Reddit), Bitchute (an alternative to YouTube)~\cite{trujilloWhatBitChuteCharacterizing2020a}, Gab, Gettr, and Parler (alternatives to Twitter)~\cite{zannettouWhatGabBastion2018, paudelLongitudinalStudyGettr2022, aliapouliosLargeOpenDataset2021}.
Previous studies uncovered radicalization within migrated communities but also showed connections to real-world events on alternative platforms~\cite{aliapouliosLargeOpenDataset2021, mcilroy2019welcome, balci2023data}.
Therefore, exploring these alternative platforms and understanding their influence is a growing focus of research efforts.

\begin{figure*}[htbp]
\centering
\includegraphics[width=0.99\linewidth]{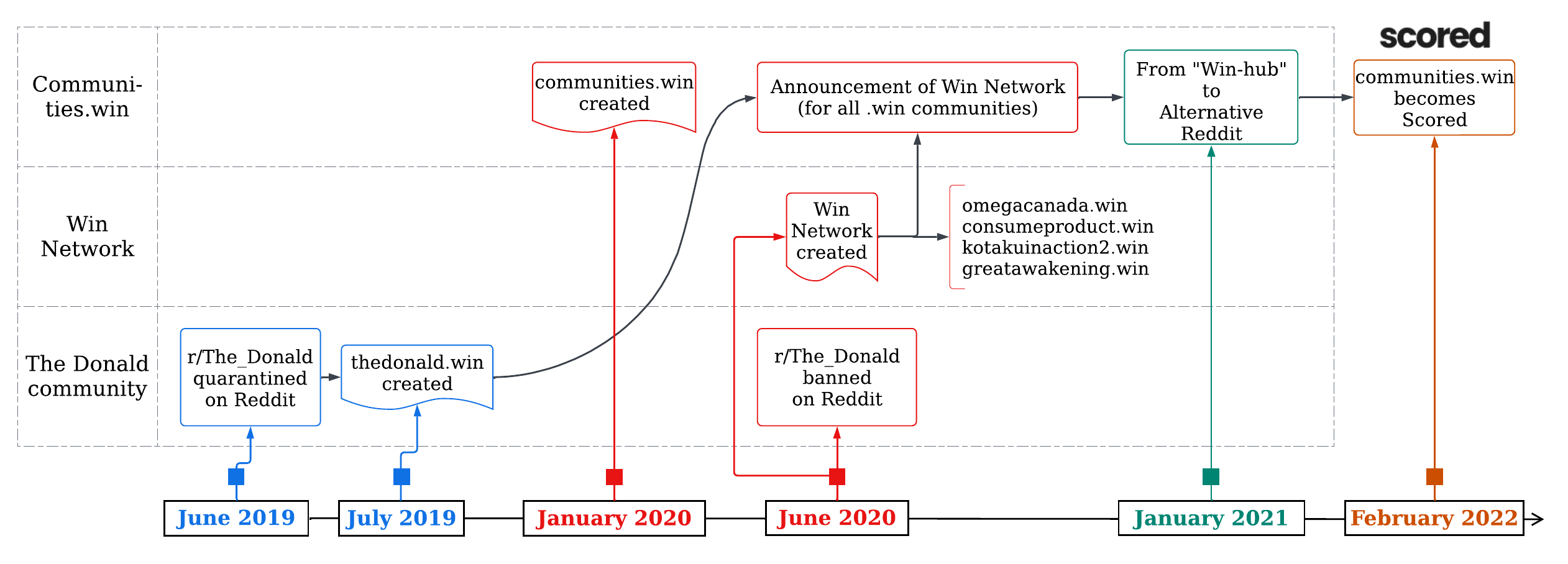}
\caption{Evolution of Scored.}
\label{fig:scored_emergence}
\end{figure*}

Scored\footnote{Scored is accessible from both \url{https://scored.co} and \url{https://communities.win}.}, an alternative to Reddit, is an emerging platform that has become a refuge for many communities banned from mainstream platforms.
We focus on collecting data from the Scored platform since its inception.
As of December 2023, Scored hosts c/TheDonald (a community for Donald Trump's supporters) and other communities (e.g., c/ConsumeProduct, c/GreatAwakening, c/FatPeopleHate) created in response to bans from mainstream platforms.
Scored remains understudied, and our work contributes to advancing the exploration of banned migrant communities on alternative platforms by providing large-scale data.

\textbf{Contributions.} In this work, we release a dataset collected from Scored.
We collect approximately 57M posts from Scored, spanning four years.
Our contributions go beyond releasing Scored's data alone; using a state-of-the-art model, we create and release sentence embeddings of the collected data.
We adhere to FAIR principles in releasing this data, discussed in a later section.

\begin{adjustwidth}{-.225in}{}
\centering
\small
\begin{tabular}{m{\baselineskip}l}
    
    \hspace*{0.2em} \includegraphics[height=0.035\textwidth]{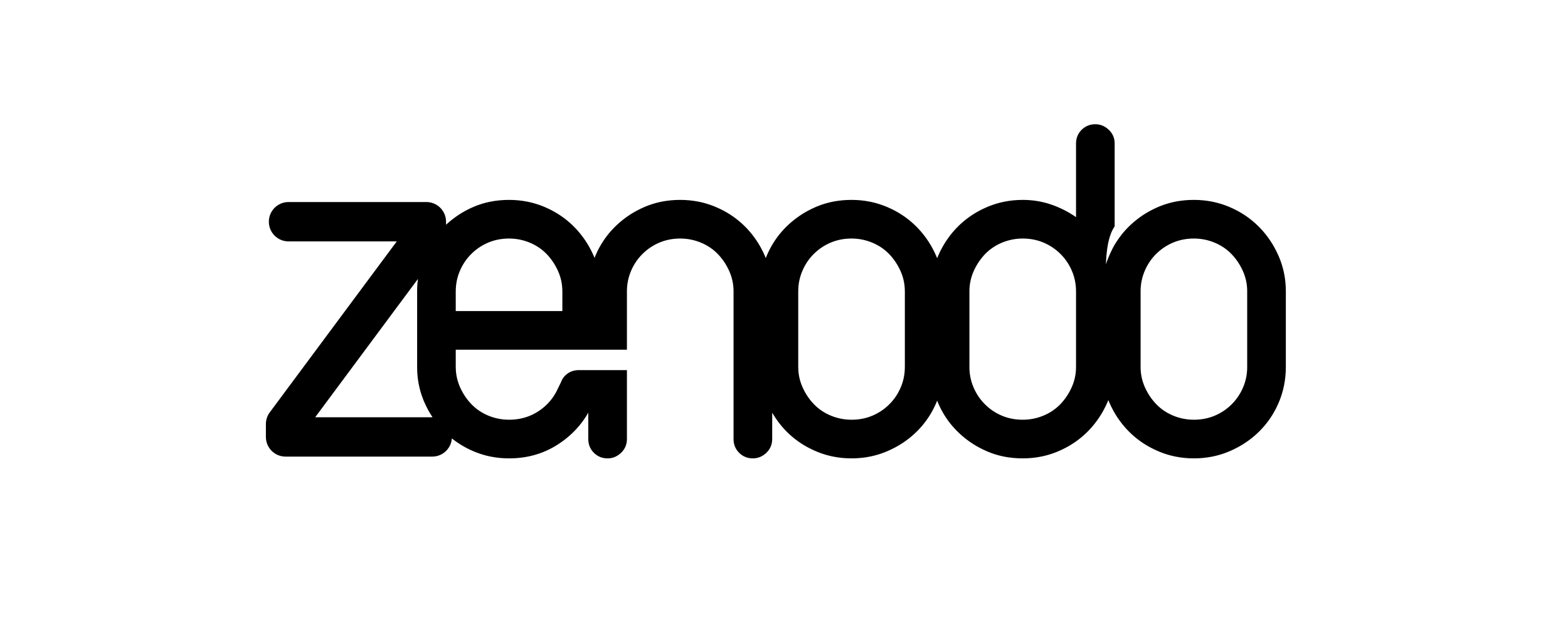}
        & 
        \hspace*{0.3em} \quad \quad \quad \color{darkgreen}{\url{zenodo.org/records/10516043}} \\ 

    \hspace*{0.8em} \includegraphics[height=0.05\textwidth]{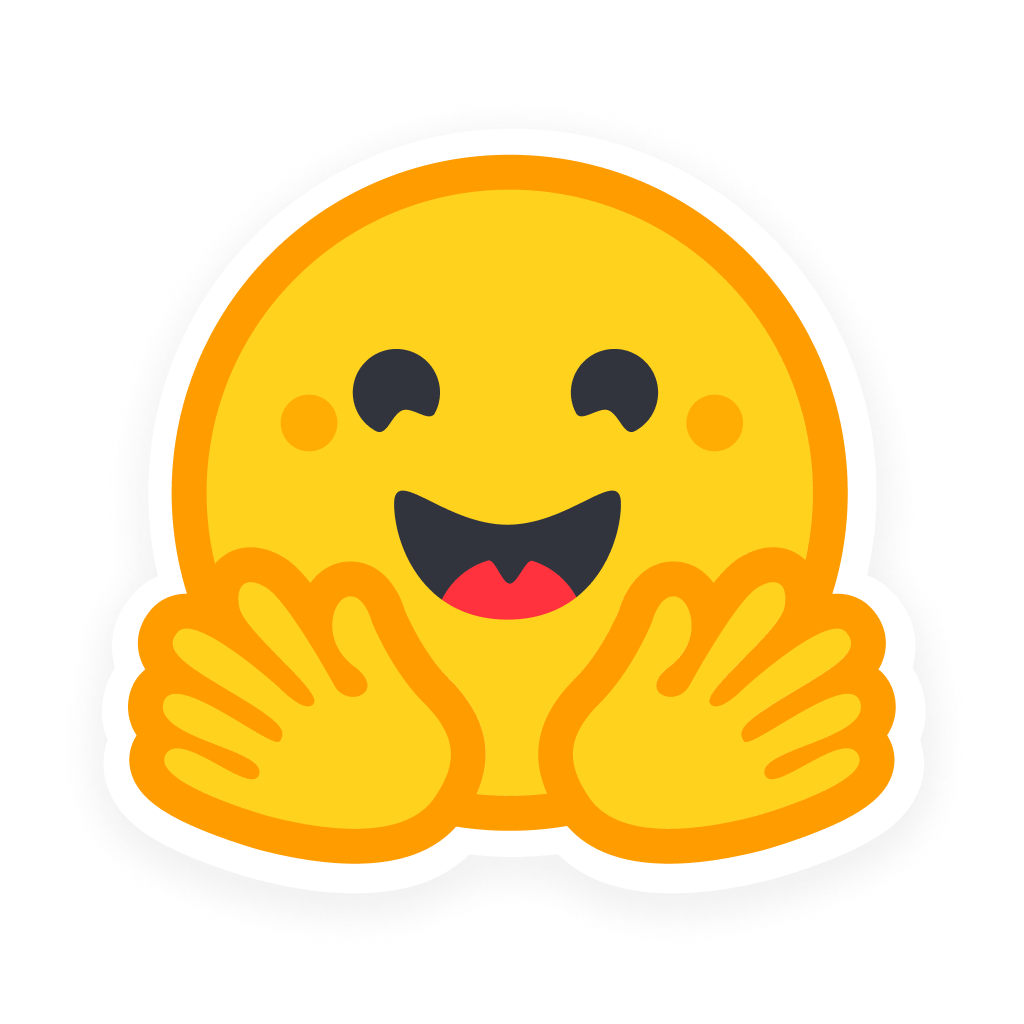} 
        & 
        \hspace*{0.3em} \quad \quad \quad \color{darkgreen}{hf.co/datasets/iDRAMALab/iDRAMA-scored-2024} \\

    \hspace*{1.1em} \includegraphics[height=0.037\textwidth]{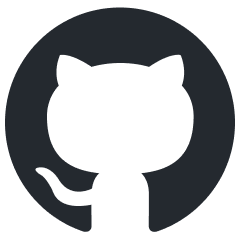}
        & 
        \hspace*{0.3em} \quad \quad \quad \color{darkgreen}{\url{github.com/idramalab/iDRAMA-scored-2024}} \\

\end{tabular}
\end{adjustwidth}

\textbf{Relevance.} With the release of this dataset, we believe that it will be valuable to the research community in various ways.
Scored emerged as an alternative platform to provide shelter to many banned communities after ``The Great Ban'' event (Reddit banned around 2,000 communities in June 2020)~\cite{cima2024great, reddit-ban, reddit-ban-verge}.
Scored hosts the far-right community, c/TheDonald, the QAnon conspiracy community, c/GreatAwakening, and more extremist ideological communities like c/ConsumeProduct, c/IP2Always.
This dataset constitutes an essential resource for researchers in understanding the migration pattern and radicalization of banned communities in online spaces.

Moreover, Scored emerged at a crucial time in U.S. history, during the U.S. Capitol riots on January 6th, 2021, an event where protestors stormed the U.S. Capitol in denial of the newly elected president `Joe Biden.'
This event fueled discussions among communities like c/TheDonald, c/GreatAwakening, and c/ConsumeProduct.
Our dataset includes the period of this event, which we believe will be helpful for researchers and journalists understanding real world events and its interplay with fringe social networks from a data-driven approach perspective.
Overall, this dataset provides a unique view to study communities' platform migration patterns and radicalization in online spaces, constituting historical political events.

\section{Background}
\label{sec:background}

As an emerging platform, Scored has a limited body of literature documenting its evolution.
In this section, we discuss the history and outlook of Scored social media platform.

\subsection{History}

In this section, we discuss Scored's history, formerly known as ``Communities.win,'' outlining the key milestones presented as a high-level overview in Figure~\ref{fig:scored_emergence}.

\textbf{Migration of r/The\_Donald Community (Jun-Jul 2019).} In June 2019, Reddit quarantined the alt-right subreddit r/The\_Donald, due to policy violations involving hate and violence~\cite{ribeiroPlatformMigrationsCompromise2021}.
Within a month of the quarantine, the community began to migrate to a self-hosted platform named ``thedonald.win.''
Records obtained from VirusTotal\footnote{All domain informations are retrieved through VirusTotal. API reference: \url{https://developers.virustotal.com/reference/overview}} show the domain (thedonald.win) being registered in July 2019.
The migration of r/The\_Donald to a self-hosted platform was somewhat unique, especially compared to other Reddit communities, e.g., r/greatawakening, that opted for Voat.co~\cite{papasavvaItQoincidenceExploratory2021}.

\textbf{Rise of Communities.win (Jun 2020).} The Communities.win domain was registered in January 2020, but no webpage references are found on the Wayback Machine until June 28, 2020.
The first post\footnote{Announcement of Win communities (June 30, 2020): \url{https://web.archive.org/web/20200630065948/https://communities.win/p/GIEblc8H/list-of-win-communities/}} on communities.win introduced a Win-network (i.e., a list of Win communities) during the same month when Reddit banned over 2,000 subreddits in June 2020.
Additionally, eight Win communities -- OmegaCanada.win, GavinMcInnes.win, ConsumeProduct.win, KotakuInAction.win, KotakuInAction2.win, WeekendGunnit.win, GreatAwakening.win, and TuckerCarlson.win -- were registered within that same month.
This suggests a possible coordinated effort and/or a migration trend, potentially influenced by a strategic move made by r/The\_Donald community in response to the Reddit ban.
During this period, Communities.win predominantly functioned as an information hub for Win network announcements.

\textbf{Community-Oriented Shift (Jan 2021).} Transforming from a central hub for Win network updates, Communities.win responded to user requests\footnote{Community request thread: \url{https://web.archive.org/web/20210123025045/https://communities.win/p/11S13qIEl9/community-request-thread/}} by hosting non-political communities (e.g., Funny and Gaming).

\textbf{Emergence of Scored (Feb 2022).} Communities.win adopted the name ``Scored'' and a new logo\footnote{Appearance of `Scored' name and logo: \url{https://web.archive.org/web/20220223204351/https://communities.win/}} in February 2022.
Simultaneously, the web page of Scored.co, which was registered in 2012, but was for sale until at least July 2019, displayed an interface identical to that of Communities.win.

\subsection{What is Scored?}

Scored describes itself as ``a platform looking to unblur the lines between entertainment and politics.''

\textbf{Users \& Content Accesibility.} Scored automatically enrolls\footnote{Automatic enrollment to the listed communities is effective as of June 2023.} new users in about ten communities, for example, Memes, Food, Technology, Animals, Sports.
Additionally, users can access posts from unjoined communities through search or direct visits.
Posts act as discussion threads, allowing users to comment and have the option to include various media types like images, videos, hyperlinks, or text.
Posts also display flair, similar to hashtags, and offer the option to be marked as Not Safe For Work (NSFW).

\textbf{Moderation.} According to Scored's content policy\footnote{Scored content-policy: \url{https://help.scored.co/knowledge-base/content-policy/}}, the platform enforces violations through warnings, temporary or permanent removal of users or privileges, and, to some extent, imposing restrictions on the community.
Scored also has one or more moderators for each community.
However, the content policy does not explicitly mention the ban of a community; instead, they provide options to mark community as unsafe for work or as political.

\textbf{Monetization.} Scored offers a pro badge to paid subscribers at USD 6.99 per month.
Pro subscribers get ad-free browsing, can give gold badges to posts or comments, and can award a one-week pro subscription to others.

\section{Dataset}
\label{sec:dataset}

In this section, we discuss the data collected from Scored, detailing the methodology used for collection and addressing potential limitations.
Furthermore, we discuss the dataset structure, data release and ethical considerations in collecting data at scale.

\begin{table}[t]
\centering
\small
\begin{tabular}{lrrr}
\toprule      
 & \textbf{Count} & \textbf{\#Users} & \textbf{\#Communities} \\
\midrule
Submissions & \numprint{6293980} & \numprint{131692} & \numprint{975} \\
Comments & \numprint{50521604} & \numprint{194084} & \numprint{686}  \\
\bottomrule
\\[-7pt]
\textbf{Total Posts} & \textbf{\numprint{56815584}} & \textbf{\numprint{226765}} & \textbf{\numprint{976}} \\
\bottomrule
\end{tabular}
\caption{Overview of dataset}
\label{tab:data_overview}
\end{table}

\subsection{Data Collection}

Using a custom crawler, we collect Scored data ranging from January 1, 2020, to December 31, 2023.
This yields a dataset of 50.52M comments and 6.29M posts (see Table~\ref{tab:data_overview}).
Table~\ref{tab:data_overview} also provides a comprehensive overview of the count of unique authors contributing at least once during this timeframe and the number of communities within our dataset.
Scored's posts are functionally equivalent to Reddit's submissions.
Throughout the rest of the paper, we use Reddit terminology as readers may be more familiar with it.
Therefore, we refer to Scored posts as a combination of Scored submissions and comments.

\textbf{Limitations.} It is important to highlight that Scored employs moderation practices leading to users' posts removal.
During our data collection, we observe instances where some posts lack author or comment/submission content, indicating their deletion before our crawler collect it.
Another limitation is associated with the attributes \textit{score}, \textit{score\_up}, and \textit{score\_down} (see Section Dataset Structure).
These metrics are collected only once through our crawler during the data collection phase without additional updates to the attributes.
Hence, dataset users should be careful while utilizing these metrics for analytical purposes.

\textbf{Ethical Considerations.} In this work, we collect and analyze publicly available data without engaging with users directly.
As a result, our institution's Institutional Review Board (IRB) does not classify this work as human subject research.
Nevertheless, we adhere to established ethical guidelines, as outlined by Rivers and Lewis~\cite{riversEthicalResearchStandards2014}, to protect the rights and privacy of users.
In the release of this dataset, we hash the actual usernames and plan to make them available only to bona fide researchers upon request.
In this paper, we also present aggregated data and refrain from any attempts to re-identify individuals within our dataset.

\subsection{Dataset Structure} Our dataset contains various attributes collected through our custom crawler. Below, we outline the attributes available in our dataset for Scored submissions:

\begin{itemize}

    \item \textbf{Identifier:} Unique identifier associated with each submission \textit{(uuid)}.
    
    \item \textbf{Primary Attributes:} Essential information about submissions, including \textit{author}, \textit{community}, \textit{created}, and \textit{date}.
    Note that the \textit{created} attribute is in Milliseconds since Unix Epoch. We convert this into \textit{date} for ease of analysis for dataset users.

    \item \textbf{Content:} Each submission includes a \textit{title} attribute representing the submission title and a \textit{raw\_content} attribute containing the submission content.

    \item \textbf{Metadata:} Attributes providing metadata for each submission are: 1) \textit{link}: URL if the submission is a link, 2) \textit{type}: indicating whether the submission is text or a link, 3) \textit{domain}: base domain if the submission is a link, 4) \textit{tweet\_id}: associated tweet id if the submission is a Twitter link, and 5) \textit{video\_link}: associated video link if the submission is a video.

    \item \textbf{Flairs:} Similar to Reddit submission flairs, which is a way to tag a submission with a certain keywords, allowing end-users to know what the submission is about. Our dataset includes two attributes: \textit{post\_flair\_text} and \textit{post\_flair\_class.} Examples of \textit{post\_flair\_text} include HIGH ENERGY, SLEEPY JOE, MEME ARMY, and more.

    \item \textbf{Scores:} Metrics providing information about the scores associated with sampled submissions, including \textit{score\_up}, \textit{score\_down}, and \textit{score}. Score attributes reflect user feedback, similar to voting, on submissions or comments. We collect these metrics only once during data collection, with no further updates.

    \item \textbf{Submission Specific Flags:} Attributes indicating specific flags for submissions, such as whether the submission is nsfw (not safe for work), posted by an admin, and more. The available attributes include: \textit{is\_nsfw}, \textit{is\_admin}, \textit{is\_image}, \textit{is\_video}, \textit{is\_deleted}, \textit{is\_twitter}, and \textit{is\_moderator}.

\end{itemize}

For all Scored comments, the following attributes are available: 

\begin{itemize}

    \item \textbf{Identifier:} \textit{uuid}.

    \item \textbf{Content:} \textit{raw\_content}.

    \item \textbf{Primary Attributes:} \textit{author}, \textit{created}, \textit{community}, \textit{date}.
    
    \item \textbf{Scores:} \textit{score}, \textit{score\_up}, \textit{score\_down}.

    \item \textbf{Comment Specific Flags:} \textit{is\_deleted}, \textit{is\_moderator}.

\end{itemize}

\subsection{Data Embeddings}

Using the transformer-based technique, specifically the INSTRUCTOR~\cite{su2022one} model, we generate embeddings\footnote{Embeddings are generated using INSTRUCTOR-large model; details here: \url{https://github.com/xlang-ai/instructor-embedding}} of 768 dimensions with a float 32-bit data type.
This state-of-the-art model is optimized for diverse downstream tasks, for example, classification, clustering, and semantic similarity.
We used a NVIDIA A100 GPU with 80GB of memory to generate these embeddings, running for approximately 40 hours.

\textbf{\textit{Embedding Methodology:}} We generate embeddings for all posts using attributes from our collected dataset.
Specifically, we use the \textit{raw\_content} field for comments.
However, for submissions, we use two attributes: \textit{title} and \textit{raw\_content}, and combine the content of both before generating sentence embeddings.
Then, we perform a preprocessing step, which involves removing hyperlinks from each post.
Also, we exclude all posts if their character length becomes zero after this preprocessing step.
Finally, using the INSTRUCTOR model, we generate embeddings for each post, resulting in approximately 48M embeddings.
NB: We also release processed posts with the embeddings.

\textbf{Data Release \& FAIR Principles.} The data is accessible on Zenodo and on Huggingface.
The data is freely downloadable and released in a standardized JSON format for interoperability.
We adhere to FAIR (Findable, Accessible, Interoperable, Re-usable) guidelines for data release~\cite{wilkinson2016fair}, ensuring its discoverability through the digital object identifier (DOI).
Additionally, we release the code for getting started with our dataset, which is available on our GitHub.
All the links to access resources are given in the introduction of this paper.

\section{General Characterization}

In this section, we present basic statistics regarding our dataset, highlighting users' posting activities on the platform.
For the top 15 communities, we discuss community-wise posting activity and also provide a brief background on their migration pattern across various platforms.
Finally, we identify migrant communities on Scored that were previously banned on Reddit.

\subsection{Posting Activity}
\label{sec:posting_activity}

\begin{figure*}[t]
\centering
  \begin{subfigure}[b]{0.48\textwidth}
    \includegraphics[width=\textwidth]{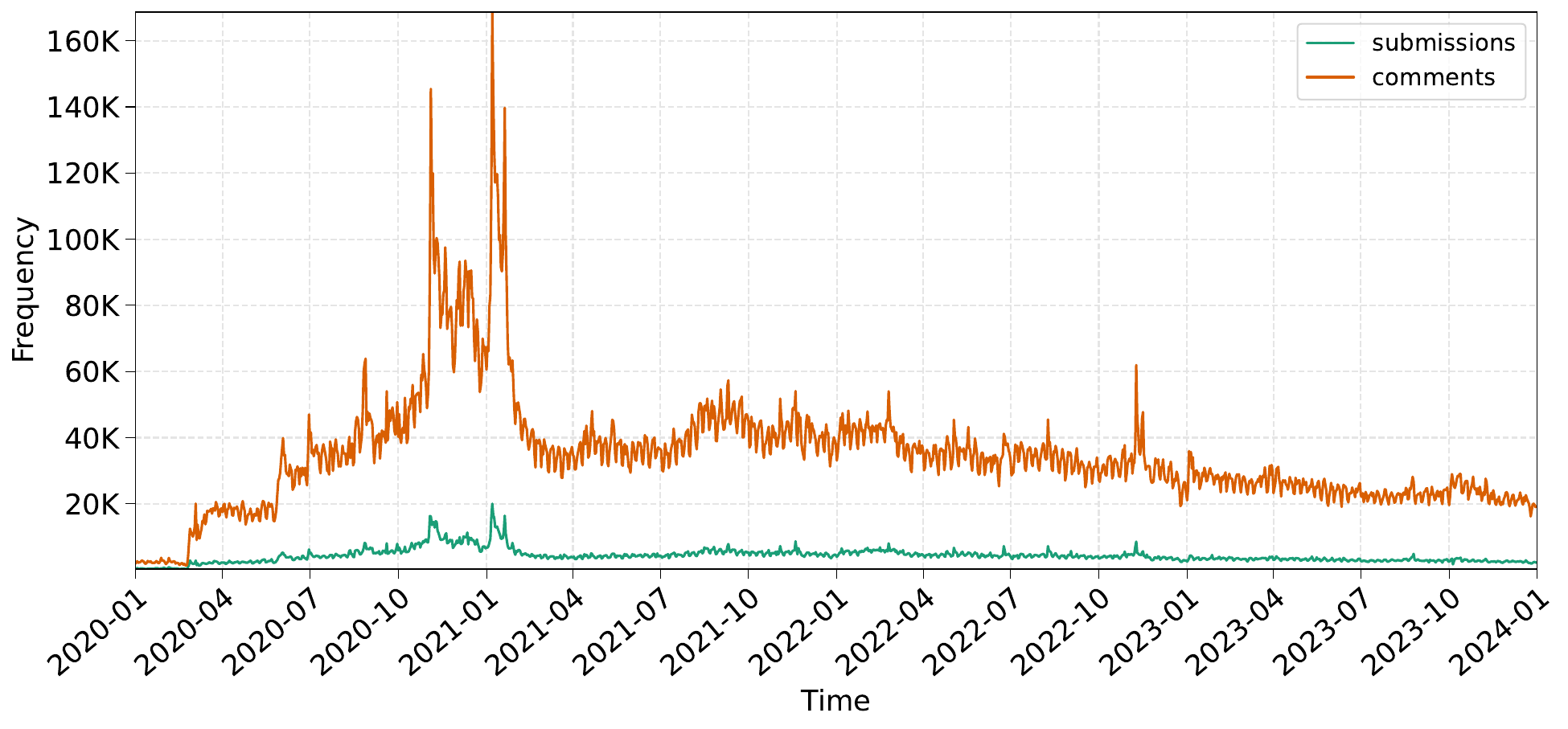}

    \caption{}
    \label{fig:posting_temporal_evolution_LEFT}
 \end{subfigure}
 \begin{subfigure}[b]{0.48\textwidth}
    \includegraphics[width=\textwidth]{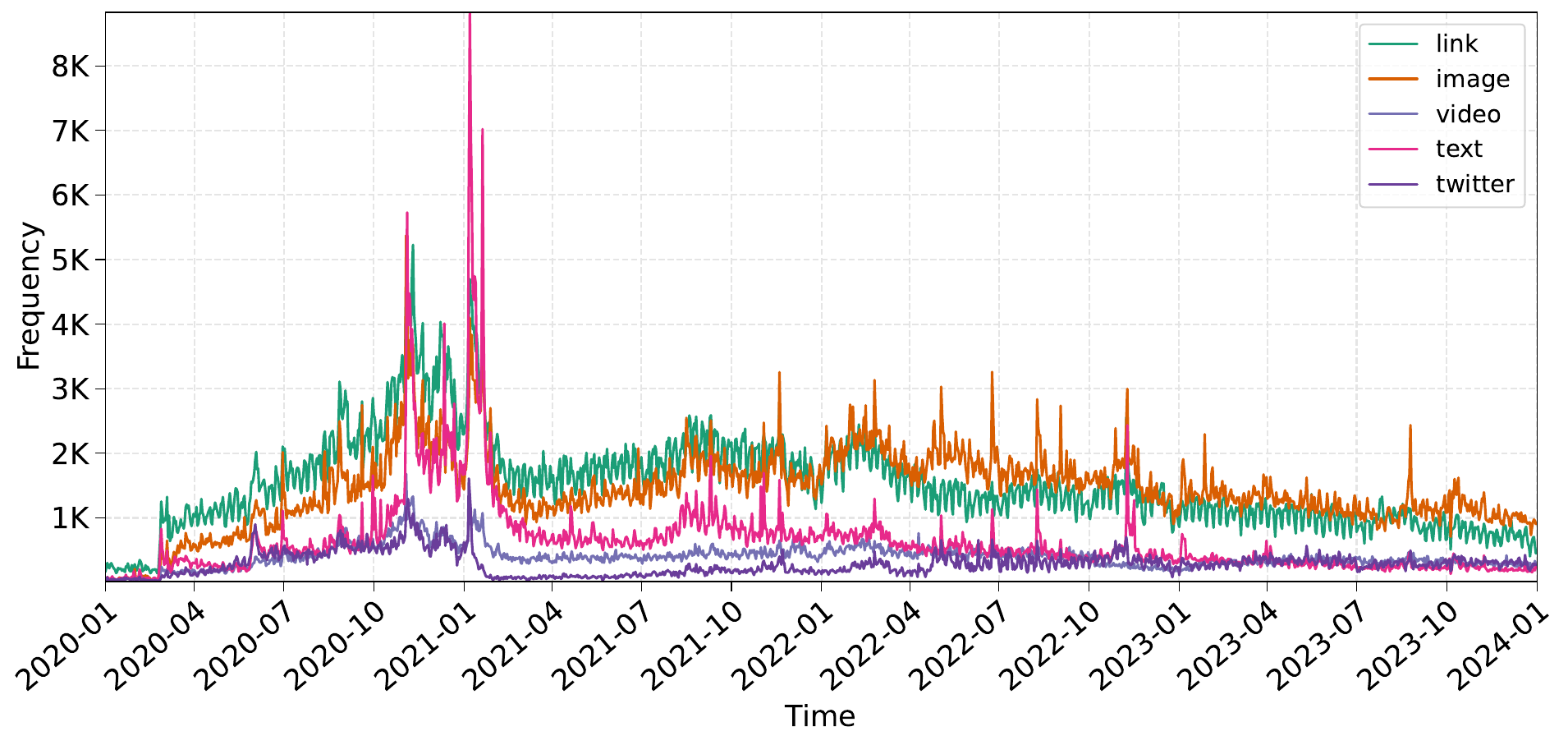}

    \caption{}
    \label{fig:posting_temporal_evolution_RIGHT}
 \end{subfigure}
\caption{Temporal evolution of daily activity in our dataset: (a) daily number of submissions/comments; and (b) daily number of submissions categorized in different types. X-axis shows daily ticks with a 3-month time interval and Y-axis shows the frequency.}
\label{fig:posting_temporal_evolution}
\end{figure*}

\begin{table}[t]
\centering
\small
\begin{tabular}{lr}
\toprule      
\textbf{Submission-types} & \textbf{Count} \\
\midrule
\textbf{Link:} Submission with hyperlinks. & \numprint{2232594} \\
\textbf{Image:} Submission includes image. & \numprint{2115019} \\
\textbf{Video:} Submission includes video. & \numprint{579150} \\
\textbf{Twitter:} Submission includes link to Twitter. & \numprint{419518} \\
\textbf{Text:} Submission with a text content, & \numprint{947724} \\
     excludes all other type of submissions. &  \\
\bottomrule
\end{tabular}
\caption{Distribution of submission types.}
\label{tab:overview_post_types}
\end{table}

\begin{figure*}[t]
\centering
\includegraphics[width=0.98\linewidth]{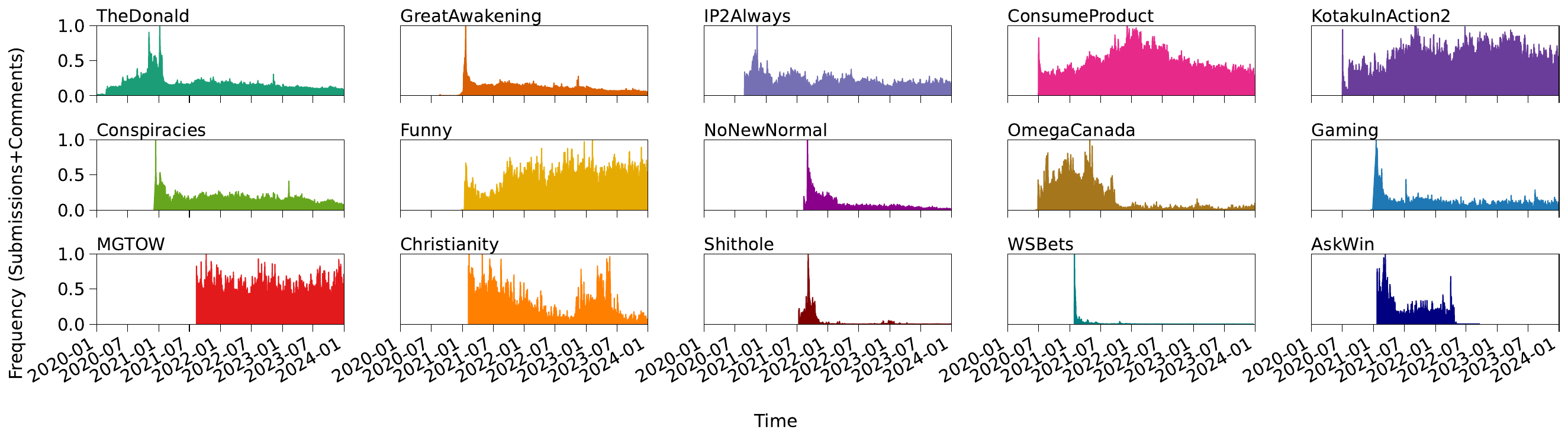}
\caption{Temporal evolution of daily activity (submissions + comments) in top 15 communities. X-axis shows daily ticks and Y-axis shows the total number of normalized posts (community-wise normalized by dividing max frequency).}
\label{fig:top15_communities_timeline}
\end{figure*}

We look at users' posting activity on Scored by creating longitudinal plot of comments and submissions over time.
Figure~\ref{fig:posting_temporal_evolution_LEFT} illustrates the longitudinal trends of submissions and comments, while Figure~\ref{fig:posting_temporal_evolution_RIGHT} shows the distribution of submission types (hyperlinks, images, videos, text, and Twitter links) over time.
We observe notable peaks in the posting activity between 2020-11-01 and 2021-01-25. 
This period includes consequential real-world political events, for example, the 2020 U.S. presidential election, the 2021 U.S. Capitol attack, and the inauguration of the 46th U.S. President, Joe Biden.

Users can post media, hyperlinks, or text on Scored, with Table~\ref{tab:overview_post_types} showing the distribution of these submission types.
Approximately 35\% of submissions are hyperlinks, 42\% are media (images or videos), 6\% are Twitter links and 15\% are text submissions.
A longitudinal analysis of various submission types reveal distinct patterns during different time periods.
As shown in Figure~\ref{fig:posting_temporal_evolution_RIGHT}, at the beginning of 2020, most submissions on the platform are hyperlinks.
Particularly, during the U.S. Presidential election (around November 2020), we see a sudden surge in text submissions.
Between January 1st, 2021, and January 25th, 2021, text-based submissions clearly dominate over other types, driven by discussions sparked by real-world events (e.g., 2021 U.S. Capitol attack).

We analyze community-wise post (i.e., submissions plus comments) distribution.
The community with the highest number of posts is c/TheDonald, with over 41.74M posts, while approximately 11\% of communities (107 out of 976) have only one post (mostly, a single submission).
About 75\% of communities have 68 or fewer posts, and a mere 3\% of communities (30 total) have more than 10K posts.

\subsection{Scored Communities}
\label{sec:scored_communities}

To gain insights per community, we present statistics of the top 15 communities by the total number of posts (Table~\ref{tab:top_15_communities}).
Notably, 97.02\% of the entire Scored user base actively engages within the top 15 communities.
All communities, except for c/TheDonald, emerged later in the timeline.
Moreover, c/IP2Always demonstrates a higher posting activity per user than c/GreatAwakening and c/ConsumeProduct, despite having a smaller user base, with 3.15M posts from 14K users.

Next, we provide a brief background of each community and discuss the longitudinal trend of posting activity within these communities (see Figure~\ref{fig:top15_communities_timeline}).
Due to space limitations, we only discuss the top 15 communities in depth, highlighting the most interesting details.

\textbf{\textit{c/TheDonald.}} The largest community on Scored identifies itself as a high-energy rally for supporters of the 45th U.S. President, Donald Trump.
Our data collection period begins on January 1st, 2020, allowing us to gather data almost from this community's inception after migration from Reddit.
We discussed the migration history of TheDonald community earlier in the Background section.

\textbf{\textit{c/GreatAwakening.}} Historically, the Great Awakening originated in the 18th century as a period of religious revival in American Christian history. 
In recent times, web communities have utilized the QAnon conspiracy theory to promote a new version of the Great Awakening, using phrases like ``taking the red pill'' to symbolize adopting QAnon awareness.
Reddit banned r/greatawakening in September 2018, leading to a migration to the Voat platform~\cite{papasavvaItQoincidenceExploratory2021}.
Following Voat's shutdown on December 25th, 2020~\cite{papasavva2023waiting}, this community potentially migrated to \url{greatawakening.win}, which co-exists under the ecosystem of Communities.win (now Scored).
Additionally, Papasavva and Mariconti~\cite{papasavva2023waiting} confirm that following Voat's shutdown, users primarily considered migrating to greatawakening.win and Poal, an alternate platform.
As of December 2023, the c/GreatAwakening community on Scored consists of approximately 33K users who have contributed around 6.16M posts.

\textbf{\textit{c/IP2Always.}} IP2Always community engages in watching live streams and creates posts, memes, or short clips from streamed videos. 
IP refers to ``Ice Poseidon,'' an In Real Life (IRL) streamer with a notable presence in live streaming. 

The IP2Always community was active on Reddit between September 2017 to September 2019.
Then, it migrated to SaidIt from September 2019 to August 2020.
Later, in August 2020, the community migrated to self-hosted platform, \url{IP2Always.win}.
Despite facing bans across multiple platforms, including Reddit and SaidIt between 2017 to 2020, c/IP2Always\footnote{As of April 1st, 2024, c/IP2Always is accessible through \url{https://scored.co/c/IP2Always} and \url{https://IP2alwayswins.com}.} has emerged as the 3rd most active on the Scored after migration.
The longitudinal plot of c/IP2Always shows consistent posting activity since its inception, with an earliest post in our dataset dating back to August 26th, 2020, shortly after the ban of /s/Ice\_Poseidon2 on SaidIt\footnote{Removal of IP2 sub from SaidIt: \url{https://saidit.net/s/SaidIt/comments/65k1/we_are_being_forced_by_our_server_company_to/}.}.

\textbf{\textit{c/ConsumeProduct.}} Reddit bans ConsumeProduct to remove hate speech-related communities in mid-2020~\cite{reddit-ban}.
This community has a controversial history on Reddit, associated with fascist and anti-Semitic ideologies.
Within our dataset, the earliest post of c/ConsumeProduct is June 29th, 2020, occurring shortly after the Reddit ban. 
As of December 2023, c/ConsumeProduct has approximately 18.5K users and about 2.26M posts.

\textbf{\textit{c/KotakuInAction2.}} KotakuInAction2 community identifies itself as a gaming community and actively participates on both Scored and Reddit platforms.
r/KotakuinAction is the largest Gamergate community on Reddit and has a controversial history of racism and sexism~\cite{kia-news, kia-news2}.
On Scored, this community comprises approximately 2.8\% of the Scored's user base.

\textbf{\textit{c/NoNewNormal.}} This community presents itself as a platform for skeptical discussions surrounding the ``new normal'' resulting from the COVID-19 pandemic. 
Reddit announced the ban of r/NoNewNormal on September 1st, 2021, mentioning the spread of COVID-19-related misinformation~\cite{reddit-covid-ban}. 
In our dataset, the first post (August 11th, 2021), hints at the beginning of a migration just 15 days before the Reddit ban. 
However, c/NoNewNormal shows minimal posting activity approximately after February 23rd, 2022 (see Figure~\ref{fig:top15_communities_timeline}).

\textbf{\textit{c/OmegaCanada.}} OmegaCanada has its origins in the formerly known r/metacanada community on the Reddit.
This community describes itself as a predominantly conservative community, with many members having faced arbitrary bans on other platforms for their conservative views.
Even though there were no apparent ban threats on Reddit, the community migrates from Reddit to omegacanada.win.
Our dataset includes posts dating back to June 13th, 2020, and a longitudinal plot shows very high posting activity in the beginning.
However, approximately after September 23rd, 2021, we observe a substantial decline in posting activity.

\textbf{\textit{c/MGTOW.}} MGTOW (Men Going Their Own Way) believes society is rigged against men, advocating for the abandonment of women and, at times, western society, while also displaying extreme anti-feminist and misogynistic views~\cite{ribeiro2021evolution}.
MGTOW experienced a quarantine on the Reddit platform in January 2021~\cite{ribeiro2021evolution}.
Within our dataset, the earliest posts date back to August 11th, 2021, emerging after more than six months of the quarantine period. 
Despite being a small community with only 2K users on Scored, the longitudinal plot shows consistent posting activity.

\textbf{\textit{c/Conspiracies, c/Funny, c/Gaming, c/Christianity, c/Shithole, c/WSBets, c/AskWin.}} These communities on Scored serve as hubs for specific topic discussions, as obvious from their names.

\begin{table}[t]
\centering
\small
\begin{tabular}{lrrr}
\toprule      
\textbf{Community} & \textbf{\#Posts} & \textbf{\#Users} & \textbf{Min. Date} \\
\midrule
c/TheDonald & \numprint{41745699} & \numprint{169520} & 2020-01-01 \\
c/GreatAwakening & \numprint{6161369} & \numprint{33008} & 2020-08-10 \\
c/IP2Always & \numprint{3154741} & \numprint{14001} & 2020-08-26 \\
c/ConsumeProduct & \numprint{2263060} & \numprint{18562} & 2020-06-29 \\
c/KotakuInAction2 & \numprint{747215} & \numprint{6209} & 2020-06-30 \\
c/Conspiracies & \numprint{539164} & \numprint{10304} & 2020-12-04 \\
c/Funny & \numprint{371081} & \numprint{7451} & 2020-12-27 \\
c/NoNewNormal & \numprint{322300} & \numprint{7644} & 2021-08-11 \\
c/OmegaCanada & \numprint{249316} & \numprint{3546} & 2020-06-13 \\
c/Gaming & \numprint{181469} & \numprint{5782} & 2020-12-16 \\
c/MGTOW & \numprint{175853} & \numprint{2006} & 2021-08-12 \\
c/Christianity & \numprint{124866} & \numprint{1515} & 2021-02-05 \\
c/Shithole & \numprint{98720} & \numprint{1314} & 2021-07-12 \\
c/WSBets & \numprint{66358} & \numprint{4427} & 2021-01-27 \\
c/AskWin & \numprint{39308} & \numprint{3393} & 2021-01-21 \\
\bottomrule
\\[-7pt]
\textbf{Total} & \textbf{\numprint{56240519}} & \textbf{\numprint{220013}} &  - \\
\bottomrule
\end{tabular}
\caption{Overview of the top 15 communities in our dataset.}
\label{tab:top_15_communities}
\end{table}

\subsection{Migrant Communities}

Looking at the history of Scored's top 15 communities, we see that most migrated from Reddit.
Here, we identify Scored communities that once existed on Reddit and were subsequently banned.
To do so, first we compile a list of \numprint{2248} Reddit-banned subcommunities by referencing previous research~\cite{phadke2022pathways} (NB: Reddit does not provide an official list of such communities).
Second, to confirm the banned status of these communities, we crawl each subcommunities' homepage and verify it through the presence of a Reddit's ban warning label.
Third, we apply the Ratcliff/Obershelp string-matching algorithm to check if a Scored community's name corresponds to a banned community on Reddit, assuming that community names remain consistent, followed as in prior research~\cite{aliUnderstandingEffectDeplatforming2021b}.
We find 58 migrant communities with a $0.85$ Ratcliff/Obershelp threshold, which accommodates cases where communities change their names.
For example, on Reddit, GreatAwakening was knows as r/greatawakening, and on Scored, it becomes c/GreatAwakening.
Along with dataset release, we also release compiled list of banned Reddit communities and the name of the communities that migrated on Scored, which will help to advance the research related migration movements of banned communities in online space.

\section{User Analysis}

In this section, we look at the influx of new users throughout the dataset collection period and examine the overlap of users within the top 15 communities.

\begin{figure}[t]
\centering
\includegraphics[width=0.80\linewidth]{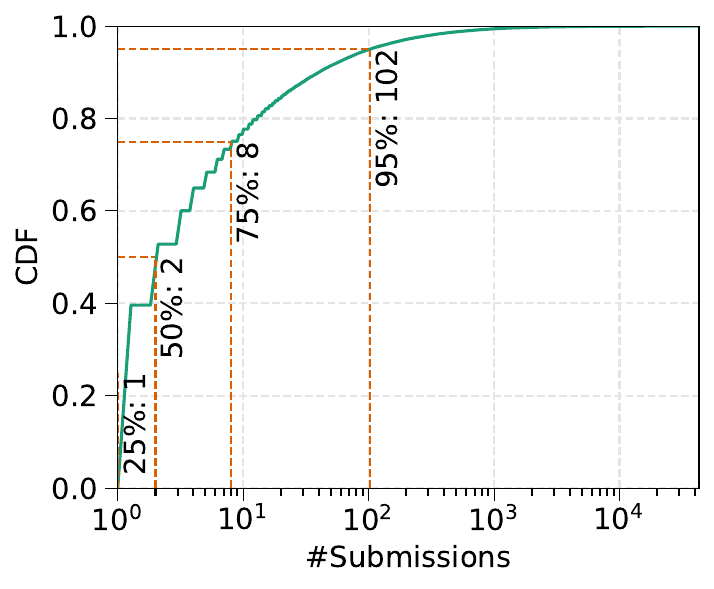}
\caption{CDF of number of submissions per user}
\label{fig:user_posts_cdf}
\end{figure}

\begin{figure}[t]
\centering
\includegraphics[width=0.80\linewidth]{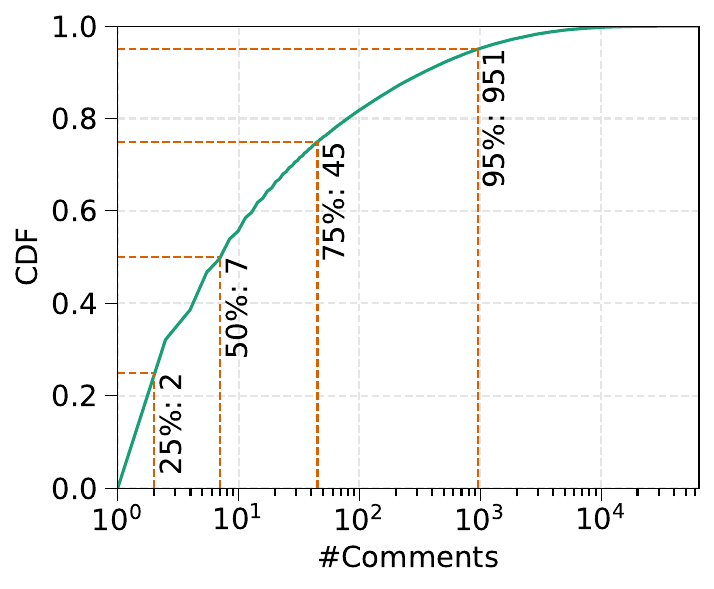}
\caption{CDF of number of comments per user}
\label{fig:user_comments_cdf}
\end{figure}

\begin{figure}[t]
\centering
\includegraphics[width=0.90\linewidth]{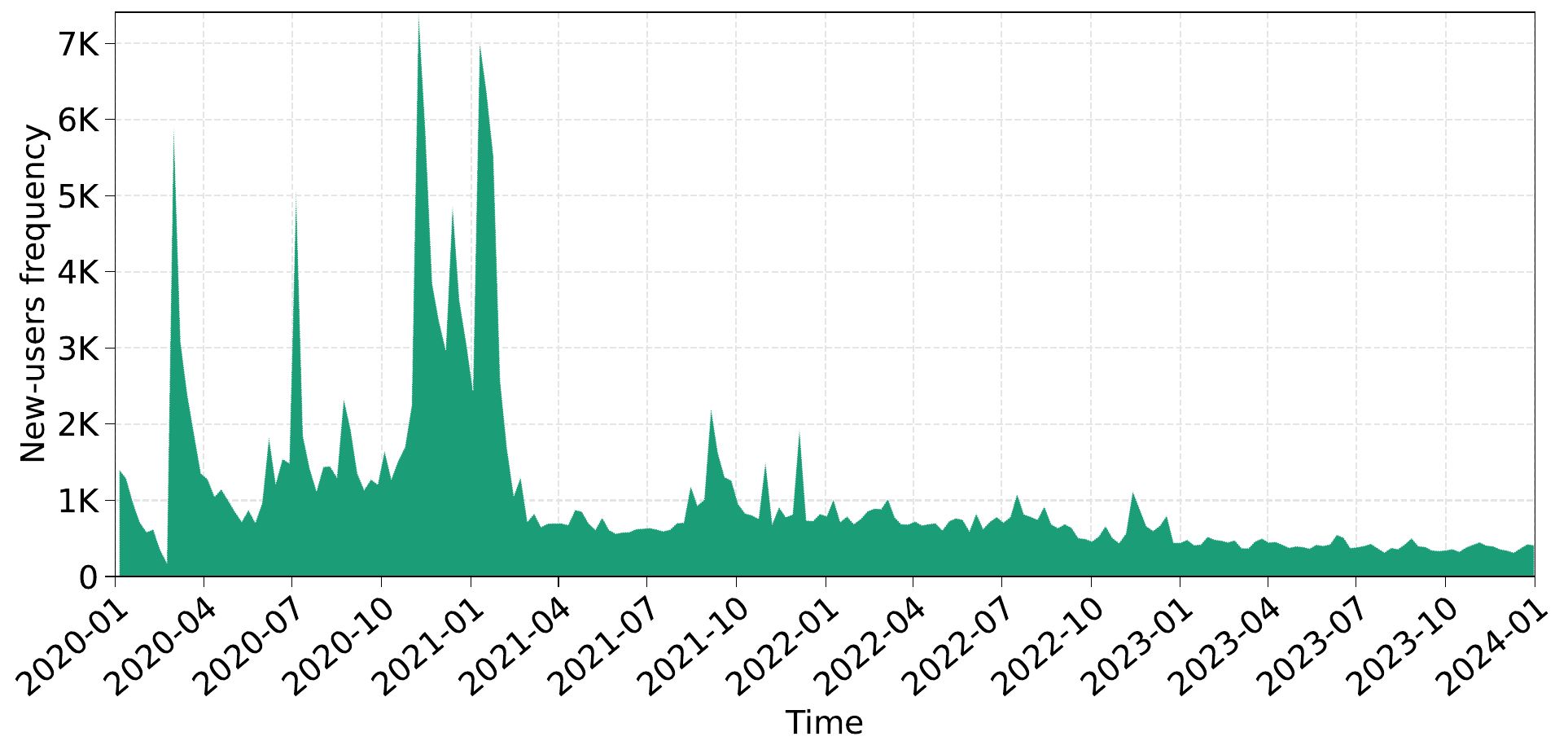}
\caption{New users over time}
\label{fig:new_users_freq}
\end{figure}

\begin{figure}[t]
\centering
\includegraphics[width=0.99\linewidth]{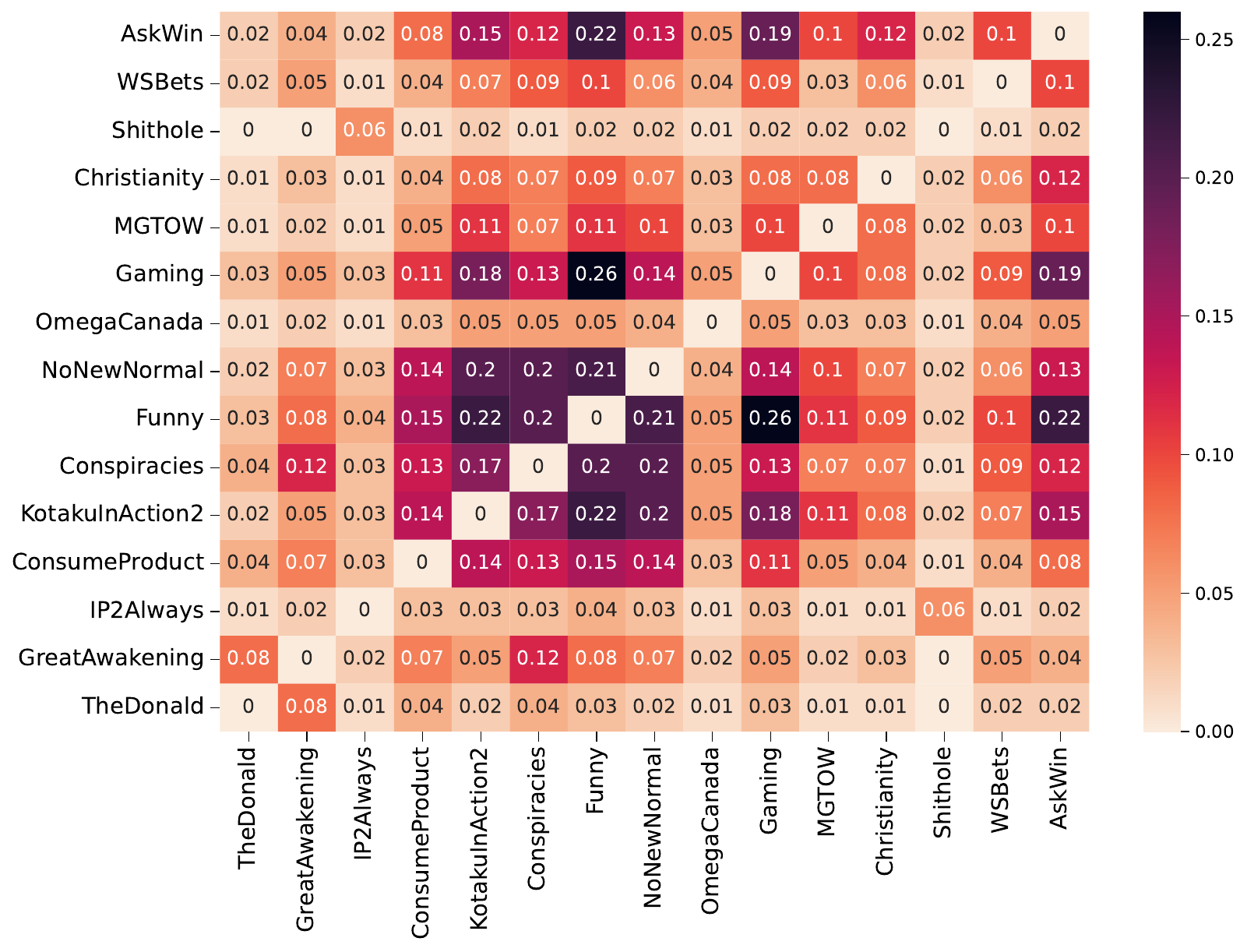}
\caption{Heatmap of user overlap between communities. Numerical values denote the Jaccard similarity between each pair of communities.}
\label{fig:user_overlap_heatmap}
\end{figure}

\textbf{User-wise Submissions/Comments.} Figure~\ref{fig:user_posts_cdf} illustrates a Cumulative Distribution Function (CDF) plot showing users' submission frequency on Scored, while Figure~\ref{fig:user_comments_cdf} displays the CDF plot of users' comment frequency.
The CDF plot shows that 95\% of Scored users have fewer than 102 submissions and fewer than 1,000 comments.
Both distributions of user participation on Scored reveal a long-tail pattern.

\textbf{Influx of New Users.} Next, we look at the influx of new users over time, as shown in Figure~\ref{fig:new_users_freq}.
Interestingly, there are points in time where a substantial number of new users joined the platform, especially during the year 2020 and early 2021.
At first, the emergence of Scored attracted new users quickly, with many communities migrating from mainstream platforms.
We also observe a surge in new users during The Great Ban event.
Another notable event that results in a surge of new users joining Scored is the U.S. election in 2020 and the U.S. Capitol insurrection.

\textbf{Inter-community User Overlap.} 
We create a heatmap to visualize user overlap between the top 15 communities, as depicted in Figure~\ref{fig:user_overlap_heatmap}.
This heatmap is constructed based on common users between pairs of communities, using the Jaccard similarity metric.
Notably, c/Conspiracies and c/GreatAwakening show a user overlap of 0.12 Jaccard value (4,511 unique users in common), while c/TheDonald and c/GreatAwakening show a user overlap of 0.08 Jaccard value (15,626 unique users in common).
Additionally, c/Funny overlaps with c/ConsumeProduct (3,410 total matches), c/KotakuInAction2 (2,421 users), c/Conspiracies (2,995 users), c/NoNewNormal (2,594 users), and c/Gaming (2,742 users).

\section{Scored Web-links Ecosystem}

\begin{table}[t]
\centering
\footnotesize
\begin{tabular}{lrrr}
\toprule
\textbf{Community} & \textbf{\#Posts} & \textbf{\#Links} & \textbf{\#Domains}\\
\midrule
c/TheDonald & \numprint{36684832} & \numprint{9756676} & \numprint{101905} \\
c/GreatAwakening & \numprint{5764811} & \numprint{963929} & \numprint{45931} \\
c/IP2Always & \numprint{2578001} & \numprint{556681} & \numprint{10743} \\
c/ConsumeProduct & \numprint{1973662} & \numprint{376741} & \numprint{17698} \\
c/KotakuInAction2 & \numprint{718374} & \numprint{110216} & \numprint{8168} \\
c/Conspiracies & \numprint{503787} & \numprint{125595} & \numprint{12300} \\
c/Funny & \numprint{358815} & \numprint{81254} & \numprint{2969} \\
c/NoNewNormal & \numprint{302031} & \numprint{54502} & \numprint{6511} \\
c/OmegaCanada & \numprint{127041} & \numprint{67812} & \numprint{5137} \\
c/Gaming & \numprint{169736} & \numprint{15115} & \numprint{2135} \\
c/MGTOW & \numprint{164717} & \numprint{16480} & \numprint{2115} \\
c/Christianity & \numprint{117373} & \numprint{24439} & \numprint{2306} \\
c/Shithole & \numprint{60564} & \numprint{15372} & \numprint{657} \\
c/WSBets & \numprint{59322} & \numprint{8323} & \numprint{1165} \\
c/AskWin & \numprint{33487} & \numprint{2474} & \numprint{856} \\
\bottomrule
\end{tabular}
\caption{Overview of extracted web-Links and unique domains for each community.}
\label{tab:communities_weblinks_overview}
\end{table}

To understand Scored's outgoing connections to the Web, we use regular expressions and extract hyperlinks from all posts.
As a result, our dataset contains 6.37M posts (i.e., submission plus comments) with one or more hyperlinks, totaling 7.39M non-unique hyperlinks.
Specifically, 4.92M hyperlinks are extracted from submissions, while 2.46M hyperlinks are extracted from comments.
Then, we extract domains from all 7.39M hyperlinks, resulting in 130K unique domains in our dataset.
Additionally, we observe that approximately 98\% of domains (approx. 127K out of 130K) have 100 or fewer links.

Table~\ref{tab:communities_weblinks_overview} provides community-wise hyperlink posting frequency across top 15 communities.
In Table~\ref{tab:communities_weblinks_overview}, \textit{\#posts} refers to the total number of posts with non-empty content (i.e., posts that have not been deleted), while \textit{\#links} indicates the number of extracted hyperlinks, and \textit{\#domains} denotes the count of unique domains from all extracted links per community.
c/TheDonald has the highest number of outgoing connections on the Web, with 9.75M hyperlinks posted, linking to 101.9K unique domains.
The community with the second-highest number of outgoing connections is c/GreatAwakening, with 45.9K domains.
c/OmegaCanada has the highest ratio (53.38\%) of extracted links to the total number of posts, followed by c/TheDonald (26.60\%) and c/Conspiracies (24.93\%).
c/MGTOW displays the lowest ratio (10.01\%) of extracted links to total posts.
Hyperlink graphs can find semantically similar websites~\cite{hanley2022no}, for instance, communities like GreatAwakening share connections with other QAnon-related conspiracy websites.
In-depth analysis of comunities' outgoing connection can help to understand communities links to the website on broder web and role of these communities to propogate information in and outside the community.

\section{Related Work}

Social networks typically enforce sanctions on users and communities for engaging in antisocial behavior, propagating hate speech, online harassment, misinformation, and conspiracy theories~\cite{trujillo2022make, ribeiroPlatformMigrationsCompromise2021, chandrasekharan2017you, mekacher2022can, jhaver2021evaluating}.
As a result, alternative platforms like Gab, Parler, Gettr, Rumble, Poal, and Truth Social have emerged in response to mainstream social media platforms.
The research community has actively curated publicly available datasets from these alternative social media platforms, emphasizing the necessity of studying them within the broader web ecosystem.
Notable contributions include datasets focusing Parler~\cite{aliapouliosLargeOpenDataset2021}, Gab~\cite{fair2019shouting}, 4chan~\cite{papasavva2020raiders}, Truth Social~\cite{gerard2023truth}, Bitchute~\cite{trujillo2022mela}, and Telegram~\cite{baumgartner2020pushshift}.
These datasets have facilitated studies analyzing online coordination in Parler leading to consequential offline events like the Storming of the U.S. Capitol~\cite{jakubik2023online}, profiling QAnon supporters on Parler~\cite{bar2023finding}, understanding temporal patterns of hate speech in Gab~\cite{mathew2020hate}, uncovering scams and conspiracy movements on Telegram~\cite{la2021uncovering}, and characterizing online incitements to harassment~\cite{aliapoulios2021large}.
The utility of these datasets extends beyond the realm of social media behavior analysis, as researchers have leveraged them to identify new cyber threats like zoombombing~\cite{ling2021first}, assess toxic behavior in online chatbots~\cite{si2022so}, and automatically detect hateful content~\cite{gonzalez2023understanding}.
Henceforth, we hope the public release and availability of Scored dataset will further improve our understanding of alternative platforms, why users migrate to these platforms, and help build better tools to identify and detect problematic content on the internet.

\section{Conclusion}

Over the past few years, there has been a tendency for many online communities, for example, r/The\_Donald and r/greatawakening, to migrate to alternative platforms following bans.
Researchers have characterized alternative platforms as problematic due to their propagation of misinformation and influence on offline events.
In this effort, we collect approximately 57M posts spanning four years (2020 to 2023) from an alternative Reddit-like social platform, Scored.
Our dataset includes over 950 communities created on Scored, from which we identified 58 communities that migrated from the Reddit ban.
Additionally, we generate and release the sentence embeddings for all the posts using a state-of-the-art model.
Furthermore, we provide a toolkit to facilitate researchers in getting started with our dataset.

Our dataset serves as a valuable resource for researchers interested in studying alternative platforms, platform migration dynamics, and fringe web communities.
In conjunction with the dataset release, we present basic statistics about banned communities and discuss the migration history of many communities across various platforms.
Leveraging this dataset can further advance research questions about web communities, e.g., do these migrant communities exploit Scored to discuss conspiracies or misinformation? what was the role played by these communities in disseminating information during political events like the 2020 U.S. election and the U.S. Capitol insurrection? And how does the web-link ecosystem of migrant communities look on the broader web?
Additionally, researchers can measure changes in overall toxicity or the spread of hate speech and misinformation over time by investigating pre- and post-migration activity and discussions.
Moreover, generated embeddings are useful for seamless topic generation or clustering techniques, or training algorithms in natural language processing, ultimately aiding in characterizing these fringe communities.

We hope that this dataset will help in understanding web communities' migration dynamics, their role in disseminating misinformation, conspiracies, or narratives, and the development of models capable of detecting hate speech propagated in these communities.

\section{Acknowledgments}

This material is based upon work supported by the National Science Foundation under Grant No. CNS-2247868 and CNS-2247867.

%\bibliography{references}

\end{document}